\documentclass[final,numberedheadings,sort&compress]{aipproc}

\layoutstyle{8x11double}
\usepackage{jab}

\newcommand{\gae}{\lower 2pt \hbox{$\, \buildrel {\scriptstyle >}\over {\scriptstyle\sim}\,$}}
    
\begin{document}

\title{Kinetic Scale Density Fluctuations in the Solar Wind}


\author{C. H. K. Chen}{address={Space Sciences Laboratory, University of California, Berkeley, California 94720, USA},email={chen@ssl.berkeley.edu}}
\author{G. G. Howes}{address={Department of Physics and Astronomy, University of Iowa, Iowa City, Iowa 52242, USA}}
\author{J. W. Bonnell}{address={Space Sciences Laboratory, University of California, Berkeley, California 94720, USA}}
\author{F. S. Mozer}{address={Space Sciences Laboratory, University of California, Berkeley, California 94720, USA}}
\author{K. G. Klein}{address={Department of Physics and Astronomy, University of Iowa, Iowa City, Iowa 52242, USA}}
\author{S. D. Bale}{address={Space Sciences Laboratory, University of California, Berkeley, California 94720, USA},altaddress={Physics Department, University of California, Berkeley, California 94720, USA}}

\begin{abstract}
We motivate the importance of studying kinetic scale turbulence for understanding the macroscopic properties of the heliosphere, such as the heating of the solar wind. We then discuss the technique by which kinetic scale density fluctuations can be measured using the spacecraft potential, including a calculation of the timescale for the spacecraft potential to react to the density changes. Finally, we compare the shape of the density spectrum at ion scales to theoretical predictions based on a cascade model for kinetic turbulence. We conclude that the shape of the spectrum, including the ion scale flattening, can be captured by the sum of passive density fluctuations at large scales and kinetic Alfv\'en wave turbulence at small scales.
\end{abstract}

\maketitle

\section{Introduction}

The solar wind contains fluctuations at a broad range of scales: from large scale solar cycle variations down to small scale turbulence at plasma kinetic scales. While studying kinetic plasma turbulence is of intrinsic interest, it is at these scales where plasma heating is thought to occur, so determining the nature of this turbulence is also important for understanding the macroscopic properties of the heliosphere.

\begin{figure}
\includegraphics[width=\columnwidth]{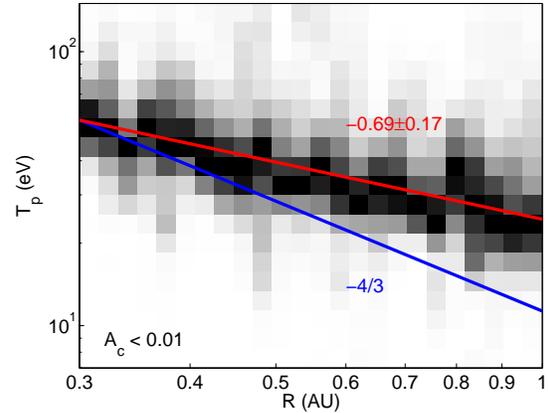}
\caption{\label{fig:tempdistance}Variation of proton temperature with heliocentric distance for collisionally young ($A_\mathrm{c}<0.01$) solar wind. The darkness represents the number of points in each bin, normalized to the maximum number for each $R$. The best fit line (red) to the peaks of a Gaussian fit at each $R$ is significantly shallower than for isotropic adiabatic expansion (blue line).}
\end{figure}

For example, it is well known that the fast solar wind proton temperature does not vary with radial distance $R$ as expected for isotropic adiabatic expansion. Fig.~\ref{fig:tempdistance} shows the radial variation of proton temperature measured by Helios Plasma Experiment \citep{schwenn75}. The plot contains only data with a small collisional age $A_\mathrm{c}<0.01$, where $A_\mathrm{c}$ is the ratio of solar wind transit time to proton collision time, e.g., \citep{bale09}. This selects the fast, hot, low density wind, i.e., the purest examples of collisionless ``fast wind.'' The radial power law is --0.69 $\pm$ 0.17, which is significantly shallower than for isotropic adiabatic expansion, for which the power law is --4/3 for an adiabatic index of $\gamma=$ 5/3 \citep{barnes74}. This non-adiabatic expansion is well known both inside \citep{marsch82a,freeman88,cranmer09} and outside \citep{mihalov78,gazis82,richardson95,cranmer09} 1 AU. In a collisionless plasma, one should ideally consider the parallel and perpendicular temperatures separately, and they also display non-adiabatic behavior \citep{marsch82a,matteini11,hellinger11}.

Dissipation of plasma turbulence is a prime candidate for the additional heating required for the non-adiabatic radial temperature profiles in the inner heliosphere \cite{matthaeus99a} (in the outer heliosphere, heating from pickup ion generated waves is thought to dominate \citep{williams95}). In order to understand solar wind heating, therefore, we need to know how the turbulence dissipates at kinetic scales. While there has been an increasing number of measurements of turbulence in this range, there is currently still disagreement about its nature (see, e.g., \citep{chen10b,salem12,chen12a} and references therein).

Recently, \citet{chen12a} measured the density fluctuation spectrum of solar wind turbulence between the ion and electron kinetic scales, finding a spectral index of $-2.75\pm0.06$. Here, we investigate this topic further by examining the nature of the flattening of the density spectrum at the ion scales.

\section{Spacecraft Charging}

The density spectrum in \citet{chen12a} was measured using the spacecraft potential \citep{bonnell08} of ARTEMIS \citep{angelopoulos11} as a proxy for density \cite{pedersen95}. This technique requires the spacecraft potential to react quickly enough in response to plasma density changes. In this section, we describe the spacecraft charging, including a calculation of the charging timescale. 

The charge on a spacecraft is given by $Q=CV_\mathrm{sc}$, where $C$ is the spacecraft capacitance and $V_\mathrm{sc}$ is the potential of the spacecraft with respect to the plasma. The time dependence, therefore, is given by 
\begin{equation}
\label{eq:cap}
dV_\mathrm{sc}/dt=I_\mathrm{t}/C,
\end{equation}
where $I_\mathrm{t}=dQ/dt$ is the total current to the spacecraft. In typical sunlit conditions, the dominant contributions to $I_\mathrm{t}$ are the flow of electrons from the plasma to the spacecraft from thermal motions $I_\mathrm{pl}=-n_\mathrm{e}eA\sqrt{k_\mathrm{B}T_\mathrm{e}/(2\pi m_\mathrm{e})}$ \citep{mott-smith26} ($n_\mathrm{e}$ = electron density, $e$ = magnitude of electron charge, $A$ = spacecraft surface area, $T_\mathrm{e}$ = electron temperature, $m_\mathrm{e}$ = electron mass), and the flow of photoelectrons from the spacecraft to the plasma $I_\mathrm{pe}e^{-V_\mathrm{sc}/T_\mathrm{pe}}$, where $I_\mathrm{pe}$ is the photoelectron current at $V_\mathrm{sc}=0$ and $T_\mathrm{pe}$ is the typical photoelectron energy in eV. We ignore higher order effects, e.g., focusing of the thermal electrons, ion currents, probe bias currents, etc. The total current is
\begin{equation}
\label{eq:current}
I_\mathrm{t}=I_\mathrm{pl} + I_\mathrm{pe}e^{-V_\mathrm{sc}/T_\mathrm{pe}}.
\end{equation}

We now examine the spacecraft potential reaction to density fluctuations. Separating the time varying quantities into mean and fluctuating parts, $I_\mathrm{pl}=I_\mathrm{pl0}+\delta I_\mathrm{pl}$ and $V_\mathrm{sc}=V_\mathrm{sc0}+\delta V_\mathrm{sc}$, and inserting Eq.~\ref{eq:current} into Eq.~\ref{eq:cap} we obtain
\begin{equation}
\label{eq:timedep}
\frac{d\delta V_\mathrm{sc}}{dt}=\frac{1}{C}\left[ I_\mathrm{pl0}+\delta I_\mathrm{pl} +I_\mathrm{pe}e^{-(V_\mathrm{sc0}+\delta V_\mathrm{sc})/T_\mathrm{pe}}\right].
\end{equation}
Subtracting the equilibrium equation, in which the currents are balanced ($0=I_\mathrm{pl0}+I_\mathrm{pe}e^{-V_\mathrm{sc0}/T_\mathrm{pe}}$), and assuming $\delta V_\mathrm{sc} \ll T_\mathrm{pe}$, Eq.~\ref{eq:timedep} becomes
\begin{equation}
\label{eq:timedep2}
\frac{d\delta V_\mathrm{sc}}{dt}=\frac{1}{C}\left[\delta I_\mathrm{pl}+\frac{I_\mathrm{pl0}}{T_\mathrm{pe}}\delta V_\mathrm{sc}\right].
\end{equation}

Eq.~\ref{eq:timedep2} is a first order linear differential equation, which can be solved to find the time dependence of $\delta V_\mathrm{sc}$ given an instantaneous change in plasma current $\delta I_\mathrm{pl}$ caused by a change in the plasma density $\delta n_\mathrm{e}$ (note that in the solar wind, temperature fluctuations can be neglected \citep{pedersen95,escoubet97b}). The solution to the equation is
\begin{equation}
\delta V_\mathrm{sc}(t)=\frac{\delta I_\mathrm{pl}\tau}{C}\left(e^{-t/\tau}-1\right),
\end{equation}
where $\tau=-CT_\mathrm{pe}/I_\mathrm{pl0}$ (note that $I_\mathrm{pl0}$ is negative so $\tau$ is positive). This describes exponential relaxation of the spacecraft potential to the new equilibrium $V_\mathrm{sc0}+T_\mathrm{pe}\delta I_\mathrm{pl}/I_\mathrm{pl0}$ with time constant $\tau$. An increase in electrons flowing to the spacecraft will result in an equilibrium with a smaller spacecraft potential.

We can now make an order of magnitude estimate of $\tau$ for ARTEMIS in the solar wind. Approximating the spacecraft by a conducting sphere of radius $L\approx0.6$ m (which gives the same surface area as the spacecraft dimensions 0.8 $\times$ 0.8 $\times$ 1 m \citep{harvey08}) gives a capacitance $C=4\pi\epsilon_0L\approx66$ pF. The characteristic photoelectron energy is $T_\mathrm{pe}\approx$ 1.5 eV \cite{grard73}. The surface area is $A\approx$ 4.5 m$^2$, $n_\mathrm{e}\approx10$ cm$^{-3}$ and $T_\mathrm{e}\approx$ 10 eV, giving $I_\mathrm{pl0}\approx$ --3.8 $\mathrm{\mu}$A. These parameters give a charging time of $\tau\approx26$ $\mathrm{\mu}$s, corresponding to a frequency $f_\mathrm{c}=1/(2\pi\tau)\approx$ 6 kHz. Since $C\propto L$, $I_\mathrm{pl0}\propto A\propto L^2$ and $T_\mathrm{pe}$ is independent of $L$, the charging time depends on spacecraft radius as $\tau\propto 1/L$, so larger spacecraft will charge quicker due to their greater surface area to collect extra charge. A plasma with higher density and temperature, having a larger return current, will also lead to faster charging.

\begin{figure}
\includegraphics[width=\columnwidth]{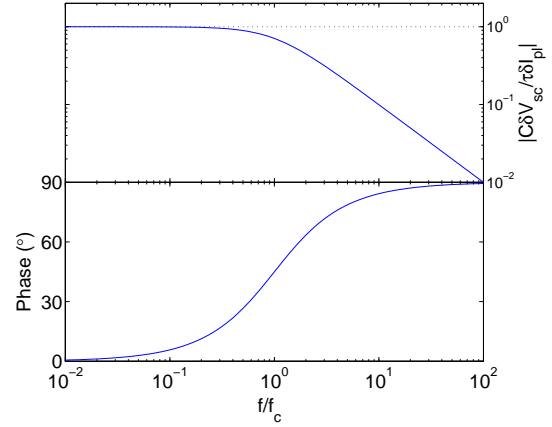}
\caption{\label{fig:response}Magnitude and phase of frequency dependent response of spacecraft potential fluctuations ($\delta V_\mathrm{sc}$) to fluctuations of thermal electron current to the spacecraft ($\delta I_\mathrm{pl}$).}
\end{figure}

Alternatively, the frequency dependence of the spacecraft potential response can be determined by substituting $\delta V_\mathrm{sc}=\delta\tilde{V}_\mathrm{sc}e^{2\pi ift}$ and $\delta I_\mathrm{pl}=\delta\tilde{I}_\mathrm{pl}e^{2\pi ift}$ into Eq.~\ref{eq:timedep2}:
\begin{equation}
\delta\tilde{V}_\mathrm{sc}=\frac{\tau/C}{1-if/f_\mathrm{c}}\delta\tilde{I}_\mathrm{pl}.
\end{equation}
The magnitude and phase of this response curve is plotted in Fig.~\ref{fig:response}. For frequencies $f\ll f_\mathrm{c}$, the potential fluctuations are linearly proportional to and in phase with the current fluctuations and do not depend on frequency, therefore the density fluctuation spectrum can be well measured by a simple calibration curve, as in \citep{chen12a}. For frequencies $f\gae f_\mathrm{c}$, the response is frequency dependent and while the density fluctuations can still be inferred (as long as the potential fluctuations are measurable), a correction for the response curve would be required.

Finally, we note that this derivation requires $\delta V_\mathrm{sc}\ll T_\mathrm{pe}$. For the frequencies $f>10^{-3}$ Hz considered here, $\delta V_\mathrm{sc}<0.1$ V so the approximation is well satisfied; for larger amplitudes, e.g., at shock crossings \citep{bale03}, the response will differ. To conclude this section, the density fluctuation spectrum of solar wind turbulence for $f\ll$ 6 kHz can be well measured using the spacecraft potential.

\section{Ion Scale Flattening}

\begin{figure}
\includegraphics[width=\columnwidth]{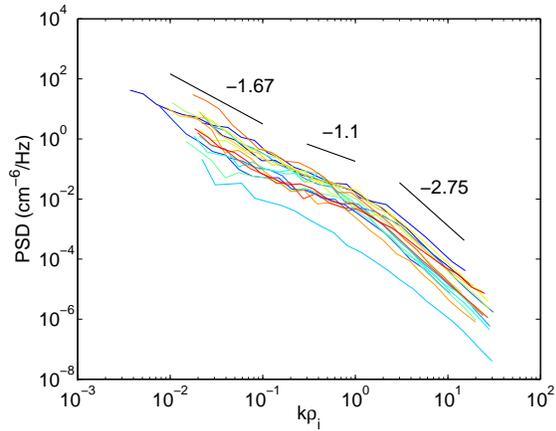}
\caption{\label{fig:allspectra}Density fluctuation spectra of 17 solar wind intervals normalized in scale to the proton gyroradius $\rho_\mathrm{i}$.}
\end{figure}

Fig.~\ref{fig:allspectra} shows all 17 of the density spectra discussed in \citet{chen12a}, with frequencies converted to wavenumber $k$ under Taylor's hypothesis and normalized to the average proton gyroradius of each interval $\rho_\mathrm{i}$. Frequencies greater than 15 Hz and less than 5 times the inverse interval length have been excluded for reliability. The measured spectral index of --2.75 for $3 <k\rho_\mathrm{i}<15$ \citep{chen12a} is marked, as well as a --5/3 inertial range spectral index \cite{chen11b}. A slight flattening of the spectrum can be seen in between, marked with a --1.1 slope to guide the eye.

The flattening at $0.1<k\rho_\mathrm{i}<1$ has been observed previously, both at 1 AU \citep{unti73,neugebauer75,celnikier83,kellogg05} and in the near-Sun solar wind \citep{coles89,coles91} and has been attributed to either pressure anisotropy instabilities \citep{neugebauer78} or the increased compressive nature of kinetic Alfv\'en wave turbulence at the ion gyroscale \citep{harmon89,hollweg99,harmon05,chandran09c}. In particular, \citet{harmon05} and \citet{chandran09c} modeled the spectrum as a sum of density fluctuations passive to the Alfv\'enic turbulence, which dominate at large scales, and active density fluctuations from kinetic Alfv\'en wave (KAW) turbulence, which dominate at small scales. At the crossover point, a flattening is naturally obtained with a shape that depends on plasma parameters.

\begin{figure}
\includegraphics[width=\columnwidth]{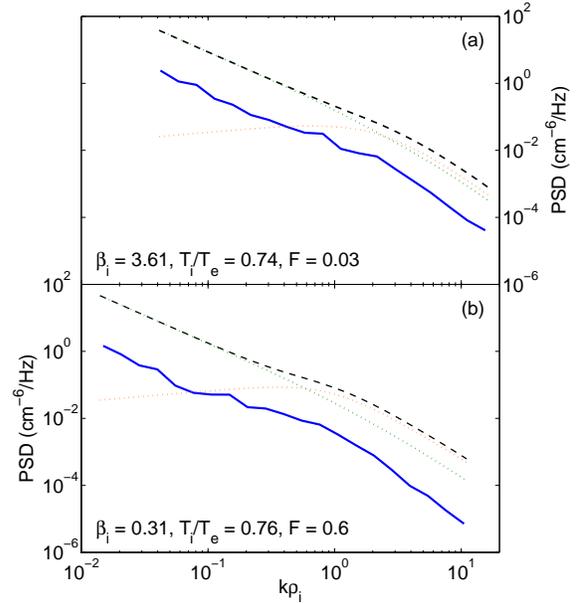}
\caption{\label{fig:beta}Density spectra (solid blue) at ion kinetic scales for (a) high $\beta_\mathrm{i}$ and (b) low $\beta_\mathrm{i}$. Theoretical spectral shapes (dashed black) with their passive (dotted green) and active (dotted orange) components are also shown.}
\end{figure}

Fig.~\ref{fig:beta} shows the density spectra for the two intervals with the highest and lowest values of proton beta, $\beta_\mathrm{i}=$ 3.61 and 0.31. The proton to electron temperature ratios are similar for these intervals, $T_\mathrm{i}/T_\mathrm{e}=$ 0.74 and 0.76, and are typical for slow solar wind. Theoretical curves, constructed using the technique of \citet{chandran09c} for the measured parameters, are also marked. The curves come from a kinetic turbulence cascade model \citep{howes11c} and consist of a passive contribution, which scales like the perpendicular magnetic field and an active contribution, calculated from the KAW eigenfunctions. To determine the relative amplitudes of these contributions, the parameter $F=[(\delta n/n_0)/(\delta v_\perp/v_\mathrm{A})]^2$ (see Eq.~1 of \citep{chandran09c}) was measured for each interval from the density and velocity spectra at frequencies $10^{-3}$ Hz $<f<5\times10^{-3}$ Hz. Note that the total power normalization of the theoretical curves is arbitrary; they have been plotted above the measured curves for clarity.

It can be seen from Fig.~\ref{fig:beta} that the shapes of the theoretical curves are in close agreement with the measurements, except, perhaps, for the slope at high frequencies. In particular, the flattening at ion scales is well captured, with the lower $\beta_\mathrm{i}$ interval showing a more prominent flattening. This is consistent with the increased compressive nature of KAWs at low $\beta_\mathrm{i}$ and is also consistent with the very prominent flattening seen in near-Sun measurements, where $\beta_\mathrm{i}$ is very low \citep{coles89,coles91}. This effect may also explain the density spectra of \citet{kellogg05}, which show more prominent ion scale flattening at low density, i.e., periods likely to have a low $\beta_\mathrm{i}$.

It is also interesting to note how $F$ varies with $\beta_\mathrm{i}$. Fig.~\ref{fig:beta} shows that the lower $\beta_\mathrm{i}$ interval has a larger $F$, meaning that the relative passive density fluctuations are larger. There are two possible explanations for this. Firstly, it is thought that the passive density spectrum above ion scales consists of kinetic slow mode like fluctuations \cite{howes12,klein12}, in which the density fluctuations are larger at lower $\beta_\mathrm{i}$. Secondly, the compressive fluctuations are less strongly damped at lower $\beta_\mathrm{i}$ \citep{barnes66,schekochihin09}.

Finally, we note that other explanations for the flattening \citep{neugebauer78,coles91,harmon05} and kinetic turbulence in general have been suggested. While there is no space here to discuss these possibilities, the model that we consider \citep{chandran09c} provides a good match to the current observations. Since the flattening is expected to be more prominent for lower $\beta_\mathrm{i}$, it should become more easily detectable in situ with future missions Solar Orbiter and Solar Probe Plus as they travel in closer to the Sun.

\begin{theacknowledgments}
This work was supported by NASA contract NNN06AA01C and grant NNX09AE41G. Helios data was obtained from NSSDC ({http://nssdc.gsfc.nasa.gov}). We acknowledge the THEMIS/ARTEMIS team and NASA contract NAS5-02099. We thank the other SW13 participants for many interesting discussions.
\end{theacknowledgments}

\bibliographystyle{aipproc_chc}
\bibliography{bibliography}

\end{document}